\documentclass[reprint,amsmath,amssymb,aps,twocolumn,]{revtex4}

\usepackage{dcolumn}
\usepackage{bm}
\usepackage{physics}
\usepackage{bbm}
\usepackage[breaklinks=true,colorlinks=true,linkcolor=blue,urlcolor=blue,citecolor=blue]{hyperref}
\usepackage{graphicx}
\usepackage{siunitx}
\usepackage[version=4]{mhchem}

\begin{document}

\title{Influence of static disorder and polaronic band formation on interfacial electron transfer in organic photovoltaic devices} 

\author{Kevin-Davis Richler$^{1,2}$ and Didier Mayou$^{1,2}$}
\affiliation{$^1$University Grenoble Alpes, Inst NEEL, F-38042 Grenoble, France\\
$^2$CNRS, Inst NEEL, F-38042 Grenoble, France}

\begin{abstract}

Understanding the interfacial charge-separation mechanism in organic photovoltaics requires, due to its high level of complexity, bridging between chemistry and physics.
To elucidate the charge separation mechanism, we present a fully quantum dynamical simulation of a generic one-dimensional Hamiltonian, which physical parameters model prototypical PCBM or \ce{C60} acceptor systems. 
We then provide microscopic evidence of the influence random static and dynamic potentials have on the interfacial charge-injection rate. 
In particular, we unveil that dynamic potentials, due to strong electron-vibration interactions, can lead to the formation of polaronic bands. 
Such dynamical potentials, when compared to random static potentials, can provide the main detrimental influence on the efficiency of the process of interfacial charge-separation. 

\end{abstract}

\pagestyle{empty}

\maketitle

\section{Introduction} 
Organic photovoltaic devices have currently attracted intense interest due to the potentially cheap production and their ability to environmentally friendly power generation \cite{li2012polymer,gunes2007conjugated,lu2015recent}.
The efficiency of these organic photovoltaic devices (OPVs) strongly depends on the charge separation process between materials that transport electrons (usually a fullerene derivative) or holes (usually a polymer).
The energy offset caused by this interface drives electrons from the donor to the acceptor, while leaving holes left behind. 
Once the charges are in separate phases, they need to overcome their mutual Coulomb attraction.
This binding energy is in the range of \SIrange{0.1}{0.5}{\electronvolt}
\cite{zhu2009charge,drori2008below,hallermann2008charge}, which is much larger than the thermal energy of about \SI{25}{\milli \electronvolt} at room temperature.
Surprisingly, the charge separation process and eventually the formation of free charges that can be extracted at the electrodes is still very efficient \cite{shirakawa1977synthesis,morel1978high}. 
In particular, experimental studies have shown, that the charge separation occurs on ultra-fast time scales in the \SIrange{10}{100}{\femto \s} range \cite{grancini2013hot,jailaubekov2013hot,gelinas2013ultrafast}.
At present, the mechanism responsible for charge separation is not well understood and still actively debated. 

An obvious question in organic materials relates to the relative importance of static and dynamic potentials and their effect on the charge separation process. 
The static (time-independent) potential reflects the electron-hole Coulomb interaction, and the spatial disorder caused by electrostatic interactions resulting from the different environment in which each molecule is placed. 
This is particularly true for organic devices made up of two different disordered materials \cite{ballantyne2010understanding,hoppe2004organic} where the spatial disorder is usually originating from the rather large permanent electric dipole of PCBM.
However, we stress here, that contrary to PCBM, molecular dynamic simulations have shown, that \ce{C60}, which has no permanent dipole, exhibits extremely limited disorder which thus can to a good extent be neglected \cite{d2016charges,tummala2015static}.
On the contrary, the dynamic potential is related to electron-vibration interactions that result in a time-dependent variation of microscopic transport parameters.
Though debated \cite{gao2014charge,de2017vibronic}, there are indications that, for charge separation, this effect could contribute \cite{falke2014coherent,song2014vibrational}.
Thus, the consideration of static and dynamic potentials seems unavoidable for a fully microscopic understanding of the charge separation mechanism in OPV devices.
However, such microscopic descriptions are computationally challenging.
To tackle these challenges, various numerical methods such as exact diagonalization \cite{wellein1997polaron,wellein1998self}, diagrammatic Monte Carlo (DMC) \cite{de1983numerical,kornilovitch1998continuous}, time-dependent density functional theory (TTDFT) \cite{rozzi2013quantum}, and other approaches \cite{jeckelmann1998density,barivsic2002variational} have been proposed. 
However, these methods are either quite expensive from a computational point of view or their application to non-translationally invariant systems remains unclear. 

To overcome these difficulties the inhomogeneous version of the dynamical mean-field theory approximation (I-DMFT) \cite{potthoff1999surface,potthoff1999metallic,potthoff1999dynamical,delange2016large,backes2016hubbard,jacob2010dynamical,turkowski2012dynamical}, a powerful non-perturbative technique for strongly interacting systems, has been introduced. 
By applying the I-DMFT approximation to a generic one-dimensional model Hamiltonian, which parameters model prototypical PCBM and \ce{C60} based acceptor systems, we provide a fully quantum dynamical simulation of the charge separation process taking static (disorder + electron-hole Coulomb interaction) and dynamic potentials (electron-vibration interaction) into consideration. 
This provides the possibility to compute the charge injection rate at the donor-acceptor interface. 
Our work provides a step forward to a long-standing challenge in OPV, thereby bridging between chemistry and physics. 
In particular we unveil here, that dynamic potentials (related to polaron formation), when compared to random static potentials, present the main detrimental lose mechanism in OPVs devices. 
Yet it can in some instances lead to an enhancement of the charge transfer process.

This paper is organized as follows. 
Section \ref{Methodology} is divided into two parts: 
(A) we introduce the generic one-dimensional Holstein based Hamiltonian to model the charge carrier dynamics across organic model interfaces 
(B) we briefly introduce the used I-DMFT approximation and comment on its general numerical aspects.
In section \ref{Results} we apply the I-DMFT approximation to study the charge carrier dynamics across donor-acceptor model interfaces.
Finally, we provide a brief conclusion and outlook in section \ref{Conclusion}.

\section{Methodology} \label{Methodology}
Inspired on the Holstein model used in \cite{bera2015impact}, we use the following generic one-dimensional Hamiltonian to describe the microscopic charge transfer process of an electron at the molecular donor-acceptor interface:
\begin{eqnarray} \label{hamiltonian_charge_separation_1d} \nonumber
H &=& \epsilon_0 \, c_{0}^{+} c_{0} + m \, c_{0}^{+} c_{1} + m \, c_{1}^{+} c_{0} +  \sum_{i=1} \epsilon_i \, c_{i}^{+} c_{i} \nonumber \\
		&-& \sum_{i=1} \frac{V}{i} \, c_{i}^{+} c_{i}  + \sum_{i=1} J \, (c_{i}^{+} c_{i+1}+ c_{i}^{+} c_{i-1} ) \nonumber \\
		&+& \sum_{i=1} \Omega \, a_{i}^{+} a_{i} + \sum_{i=1} g \, c_{i}^{+} c_{i} (a_{i}^{+} + a_{i}) 
\end{eqnarray}
where $c_{i}^{+}$ ($c_{i}$) is the creation (destruction) operator of electrons, $a_{i}^{+}$ ($a_{i}$) is the creation (destruction) operator of phonons, $\Omega$ is the relevant phonon frequency ($\hbar=1$ throughout this work), $g$ is the electron-phonon coupling strength, $m$ is the tunneling amplitude between the LUMO of the donor (site i = 0) and the first site of the LUMO of the acceptor, $V$ determines the Coulomb potential, $J$ is the electron hopping parameter, $\epsilon_0$ is the energy of the incoming electron and $\epsilon_i$ is the energy level of a molecule on the acceptor site which is taken to be a random variable (with a mean $\epsilon$ and a standard derivation $\sigma_{\epsilon}$) drawn from a rectangular distribution. 
In this model, the static potential is given explicitly by the distribution of the onsite energies $\epsilon_i$ and the Coulomb potential $V$ between the electron and hole, while the dynamic potential is caused by the electron-phonon coupling term $\sum_i g \, c_{i}^{+} c_{i} (a_{i}^{+} + a_{i})$.

At this point, we want to note, that we make the following assumptions. 
First, we do not model the hole dynamics since the effects of hole diffusion usually lead to a reduction of the Coulomb interaction only resulting in an increase of the electron-hole separation yield \cite{athanasopoulos2017efficient,aram2016charge} and since hole transfer typically occurs on time scales in the \SIrange{1}{2}{\pico\s} range which is several orders of magnitude slower than electron transfer \cite{long2014asymmetry}. 
Second, we use a one-dimensional model to describe a complex three dimensional bulk molecular heterostructure, since the main features of the Holstein polaron do not depend strongly on the dimensionality of the system \cite{ku2002dimensionality} and since we expect, that a realistic, three-dimensional nature of the system will lead to mainly quantitative changes in the quantum efficiency. 
Third, we assume a single high-frequency intramolecular mode of vibration (order of $\SI{1600}{\per \cm}$, i.e. a period of roughly \SI{20}{\femto\second}) only, although the involvement of multiple phonon modes provides additional transfer channels.   
However, in this work, we are interested in electron dynamics that concern the fast electron unbinding from the Coulomb well which occurs on time scales of \SIrange{10}{100}{\femto \s}. 
Therefore, the impact of weakly coupled low-frequency modes of vibration (order of $\lesssim \SI{80}{\per \cm}$, i.e. a period of roughly \SI{415}{\femto\second}) can be neglected, since final equilibration occurs at longer relaxation times \cite{xie2015full}.
Fourth, all computations are presented at zero temperature taking no entropic effects into consideration, since electronic and vibrational energy scales are much larger than $k_{\mathrm {B} }T$, where $k_{\mathrm {B} }$ is Boltzmann's constant and  $T$ is the temperature \cite{pelzer2016charge}, and since we expect that a change in entropy plays a diminished role in the charge separation process in one-dimensional systems \cite{gregg2011entropy}.
The driving force for the charge separation mechanism is thus entirely of quantum mechanical nature steaming from the coupling of an initially discrete state (electron at the interface) to a final state in the continuum (electron on the acceptor side surrounded by a cloud of phonon excitations).
Fifth, we note that in absence of electron-phonon interaction scattering on defects can lead to Anderson localization \cite{anderson1958absence} and it will hinder electron transfer across the interface. 
Since it is well known, that in the case of one-dimensional infinite disordered systems, any amount of disorder produces Anderson localization, we have embedded the system into an effective medium which has been computed using the coherent potential approximation (CPA).
The absence of Anderson localization within CPA \cite{Haydock_1974} will then mimic a system without localization at moderate disorder (except very close to the band-edges) throughout this work. 

The approach that we propose is the use of the single-polaron I-DMFT approximation.
The aim of I-DMFT is to fully address the relevant spatial variations of the physical properties, but still affording a good description of the physical processes of interest. 
A crucial aspect of I-DMFT is that it provides an interpolation between the non-interacting case, in which it gives the exact solution of the problem and the strong coupling limit in which is also becomes exact.
In I-DMFT the local Green's function is for one given realization of disorder and is defined as:
\begin{equation} \label{green}
G_{ii}(z)^{-1} = z - \epsilon_i - \Delta_{i}(z) - \Sigma_{i}(z) 
\end{equation}
where $\Delta_i(z)$, $\Sigma_i(z)$ are hybridization function, respectively self-energy at site i and $z = E + i \eta$ with $\eta$ being an infinitesimal small number. 
In standard I-DMFT procedures a self-consistent solutions is obtained iteratively at each z where one needs to compute repeatedly the diagonal of the inverse of a complex matrix, which dimension equals the number of lattice sites.
Using conventional linear-algebra algorithms, this problem grows cubic with the size of the system \cite{freericks2016generalized}.

We use an alternative approach to solve the I-DMFT self-consistency equations which is based on Haydock's recursion scheme applied to suitably defined Hamiltonians and, in particular, which does not require the inverse of a complex matrix. 
Instead the Hamiltonian (\ref{hamiltonian_charge_separation_1d}) is solved on the full lattice under the approximation that the electron-phonon self-energy is local depending on the frequency of the local phonon only.
Self-consistency equations are than expressed in Hilbert-space such that the recursion technique by Haydock \cite{haydock1980solid} can be used which makes this method immediately generalizable to any lattice geometry and/or disorder distribution while easily handling inhomogeneous systems of up to $10^3$ lattice sites in less than \SI{24}{\hour} (sequential computation on an Intel Xeon E5-2670 processor).
 \begin{figure}[!t] 
  \centering  
    \includegraphics{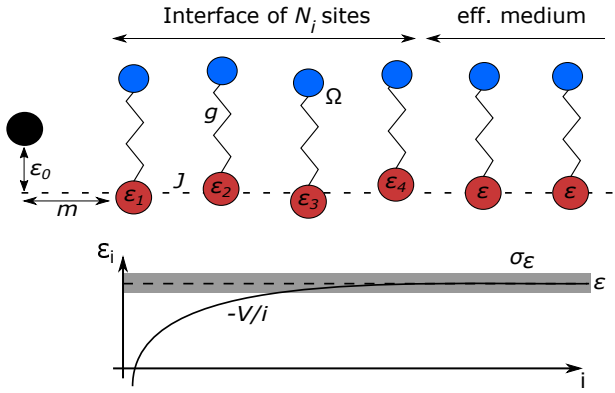}  
  \caption{Schematic representation of the Hamiltonian (\ref{hamiltonian_charge_separation_1d}) describing the electron dynamics at the donor-acceptor interface. The black curve represents the coulomb attraction from the hole (black site) and the dashed area represents the spread around the mean energy value $\epsilon$. We embedded the left most non-equivalent site $N_i$ into an effective medium which has been computed using the coherent potential approximation.} 
    \label{System_sketch_interface_1d}
 \end{figure}
  \begin{figure*}[!t]
  \centering
        \includegraphics[scale=1.0]{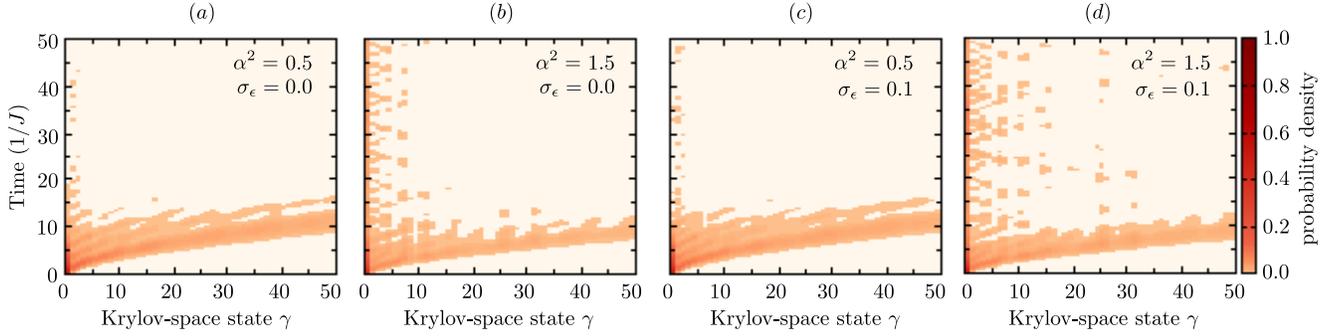}  %
    \caption{Time evolution of the probability density $n_{\gamma}(t)$ at each state $\gamma$ in Krylov-space where all calculations are shown for an incoming electron energy $\epsilon_0=-0.5$ that is taken in the band of delocalized states (time unit in $1/J \approx \SI{3.3}{\femto \s}$). 
    Panel (a) and (b) are performed for dynamic potentials only, while (c) and (d) are performed for dynamic potentials and static disorder.
    In fact, molecular dynamics simulations have shown, that contrary to PCBM fullerenes, \ce{C60} exhibits extremely limited disorder in its crystalline phase, which can thus to a good extent be neglected \cite{tummala2015static,d2016charges}.
    Panel (b) and (d) show a localization effect near the interface, which are, however, superposed with outgoing wave packets, resulting in a poor interfacial charge-transfer efficiency. 
    We did not attempt to estimate the numerical error, but it should be of the order of the fluctuations visible in the plot.} 
    \label{time_evolution}
 \end{figure*} 
An detailed explanation of the used I-DMFT approach can be found in the original work by \cite{richler2018inhomogeneous}.

Finally, let us comment on general numerical aspects of the proposed I-DMFT formalism. 
The number of possible phonon configurations is infinite, but can be restricted to a finite, sufficiently large number in actual calculations by choosing $M >> g^2/\Omega^2=\alpha^2$, where $M$ is the maximum number of phonon excitations per site. 
Further, we simulate only a finite part of the total size of the system ($N$).
We then embedded the leftmost site into an effective medium which has been computed using the coherent potential approximation (CPA). 
We thus define a number $N_i < N$ of non-equivalent lattice sites along the surface.
Choosing $N = 2000$, $M=60$ and $N_i = 300$ ensures that all results are independent of any system size characteristics, while keeping modest computational complexity. 
A pictorial representation of the proposed model is depicted in Fig. \ref{System_sketch_interface_1d}.

\section{Results} \label{Results}
In the following we express all energies in units of $J$ (energy unit in $J\approx \SI{0.2}{\electronvolt}$ and time unit in $1/J \approx \SI{3.3}{\femto \s}$) and take $m=0.5$, $V= 1.5$, $\Omega = 1.0$, $\epsilon = 0.0$ and $\sigma_{\epsilon} = 0.0, 0.1$ while keeping $g$ and $\epsilon_0$ as independent parameters.  
All are chosen such, that they are in a realistic experimental range to model the prototypical PCMB ($\sigma_{\epsilon}=0.1$) and \ce{C60} ($\sigma_{\epsilon}=0.0$) acceptor systems \cite{zheng2017charge,d2016electrostatic,antropov1993phonons,faber2011electron,castet2014charge}. 
We have checked that all results are qualitatively insensitive to different disorder configurations, a result that has been found in all tested cases throughout this work. 
Finally we note, that throughout this paper our initial state at time $t=0$ will consist of an electron at site $i=0$ having energy $\epsilon_0$ and no phonon modes excited, i.e. $\ket{\psi(t=0)} = c_0^{+} \ket{0}$ with $\ket{0}$ being the vacuum state for phonons and electrons. 

First, we present in Fig. \ref{time_evolution} the probability density of the wave-function $\ket{\mathcal{K}_{\gamma}}$ on the acceptor
\begin{equation}\label{Krylov}
n_{\gamma}(t) = \abs{\braket{\mathcal{K}_{\gamma}}{\psi(t)}}^2 
\end{equation}
where $\ket{\mathcal{K}_{\gamma}}$ is the orthonormal basis vector of the $N$-dimensional Krylov-space ($KS_N$) \cite{gutknecht2007brief}.
The Krylov-states $\ket{\mathcal{K}_{\gamma}}$, which are computed by a Lanczos-based recursion method \cite{haydock1980solid}, represent a basis of excitations of the many-body system (electron and phonon modes) that progressively spread away from the interface into the acceptor.
The time evolution of the wave function in Krylov space is then defined as $\ket{\psi(t)}=e^{-i\mathcal{H}_{KS_{N}}t} \ket{\mathcal{K}_{0}}$ with $\mathcal{H}_{KS_{N}}$ being the reduced Hamiltonian of the Krylov subspace $KS_N$. 
The time-evolution has then been determined by an exact diagonalization technique by choosing the first Krylov-space vector equal to the initial state, i.e. $\ket{\mathcal{K}_{0}} = c_0^{+} \ket{0}$ (see Appendix \ref{krylov_space} for a brief introduction to the Krylov subspace method and computation of wave-function $\ket{\mathcal{K}_{\gamma}}$).
This provides an efficient way to extract the essential character of the Hamiltonian (\ref{hamiltonian_charge_separation_1d}) while using a limited number of basis sets.
We stress here, that the size of the system (2000 states in the Krylov-space) is sufficient to prevent the wave packet to bounce back at the boundary.

We find, in the limit of sole dynamic potentials ($\alpha^2=0.5$, $\sigma_{\epsilon}=0.0$) shown in Fig. \ref{time_evolution}(a), that the probability weight near the interface is progressively decaying with time and the particle is fully delocalized.  
Upon increasing the electron-phonon interaction ($\alpha^2=1.5$, $\sigma_{\epsilon}=0.0$), the local density of states (LDOS) fragments into polaronic sub-bands (self-trapping) and where the strongly renormalized width of the polaronic sub-bands arises as the new energy scale \cite{ciuchi1997dynamical}. 
As can be seen in Fig. \ref{time_evolution}(b) self-trapping of the electron hinders, in this case, the interfacial electron transfer drastically as parts of the wave-function remain localized at the interface resulting in a poor but finite interfacial charge-transfer efficiency.
In panel (c), (d) of Fig. \ref{time_evolution} we now present the combined effect of static spatial disorder and weak ($\alpha^2=0.5$, $\sigma_{\epsilon}=0.1$), respectively strong ($\alpha^2=1.5$, $\sigma_{\epsilon}=0.1$) dynamic potentials.  
As can be seen, the in panel (c), (d) presented physical picture does not change quantitatively when compared to panel (a), respectively (b), an outcome we have found throughout this work. 

Before proceeding we stress here, that the LDOS is a key factor since its shape and, in particular, the energy distribution of electronic states are found to determine the value of practically achievable injection energies.
This comes in handy when using the proposed formalism \cite{richler2018inhomogeneous}.
In particular, we have found, that taking the incoming electron energy outside the band of delocalized states completely suppresses, as expected, interfacial charge transfer (electron becomes localized at interface) while an incoming electron energy that is taken in the band of delocalized states results in interfacial charge transfer. 
 \begin{figure}[!t]
  \centering
\includegraphics[scale=1.0]{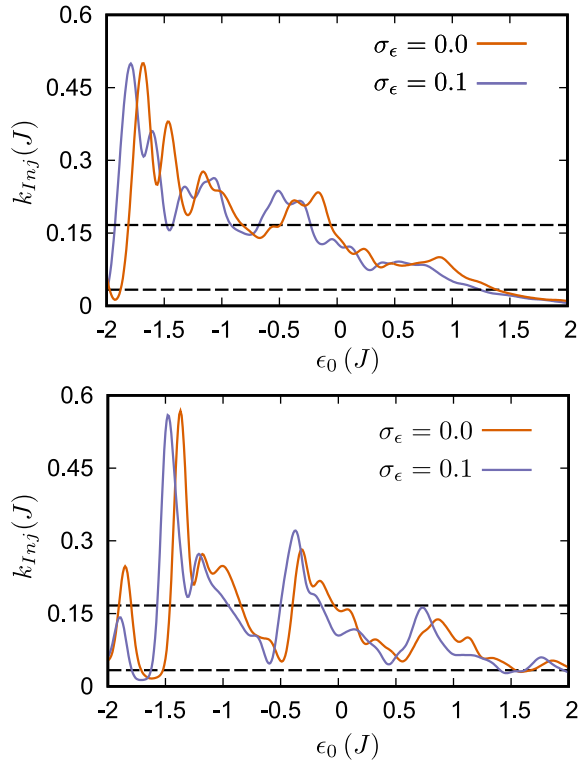}
    \caption{We present the interface transfer rate $k_{Inj}$ (rate unit in $J \approx \SI{0.3}{\per \femto \s}$) as a function of the incoming electron energy $\epsilon_0$ for $\alpha^2=0.5,1.5$ (top to bottom). 
    The dashed lines mark the range of experimentally observed recombination rates $k_{Rec}$ of order $0.033 \, J$ (\SI{0.01}{\per \femto \s}) to $0.167 \, J$ (\SI{0.05}{\per \femto \s}) \cite{grancini2013hot,jailaubekov2013hot,gelinas2013ultrafast}. 
     Transfer rates $k_{Inj}$ below $k_{Rec}$ will eventually lead to electron-hole and/or geminate recombination at the donor-acceptor interface. 
     The bottom panel shows that increasing the electron-phonon interaction $\alpha$, narrow polaronic sub-bands separate by $\Omega$ arise, an effect which hinders the range of suitable incoming electron energies $\epsilon_0$. 
 We note, that we did not attempt to estimate the numerical error, but it should be of the order of $\eta / J = 5.0 \times 10^{-2}$ due to Gaussian broadening.}
    \label{Yield}
 \end{figure} 

By making use of the by I-DMFT computed Green's function $G_{11}(z)$ and by applying Fermi's Golden rule we now quantify, as a function of the incoming electron energy $\epsilon_0$, the quantum mechanical donor-acceptor interface transfer rate
\begin{equation} \label{fermi}
k_{Inj}(\epsilon_{0}) = - 2 \, m^2 \, \Im G_{11}(\epsilon_0 + i \eta) \, 
\end{equation}
where $\Im G_{11}(z)$ is the imaginary part of the Green's function $G_{11}(z)$. 
This present a simple measure of the efficiency for charge transfer at the donor-acceptor interface, but note, the transfer rate is related, but not equal to the internal quantum efficiency which is commonly extracted from experiments. 
The interface transfer rate is graphically depicted in Fig. \ref{Yield} for incoming electron energies of $-2J \le \epsilon_0 \le 2J$,  
where this restriction is sufficient to assure the applicability of Eq. (\ref{fermi}) (weak coupling to a quasi-continuum of states).
Although, we do not model the competing electron-hole and/or geminate recombination processes, we have indicated in the same figure the range of experimentally observed recombination rate $k_{Rec}$ of order \SIrange{0.01}{0.05}{\per \femto \s} \cite{grancini2013hot,jailaubekov2013hot,gelinas2013ultrafast}.
Therefore, efficient charge transfer from the donor to the acceptor occurs when the interface transfer rate $k_{Inj}$ is large compared to the competing process of charge recombination $k_{Rec}$. 
We start considering the case of sole, but weak dynamic potentials only ($\alpha^2=0.5$, $\sigma_{\epsilon}=0.0$).
We find, that the interfacial transfer rate is essentially, with the exception of values of $\epsilon_0$ close to the low energy edge, higher than the upper experimental recombination rate for all negative values of $\epsilon_0$. 
On the contrary, this changes drastically for all positive values of $\epsilon_0$, where interfacial charge transfer is large only, when compared to extremely low experimental ranges of recombination rates. 
This stems from strong relaxation processes (phonon emission) caused by molecular vibrations which are facilitated at higher energies. 
Upon increasing the electron-phonon interaction ($\alpha^2=1.5$) multiple gaps throughout the spectrum arise and thus multiple values of $k_{Inj}$ occur that are below the experimental value of $k_{Rec}$. 
This is readily explained since upon increasing the electron-phonon interaction the LDOS fragments into sub-bands that are separated by multiples of $\Omega$.
This effect will drastically hinder the range of suitable incoming electron energies, since the incoming electron energy $\epsilon_0$ must be identical to that of the allowed unoccupied state of the acceptor material. 
Surprisingly, this picture does not change qualitatively when considering the combined effect of static disorder and dynamic potentials.
In particular, we have found that weak but static disorder does not change drastically the interfacial transfer rate for incoming electron energies of $-2J \le \epsilon_0 \le 2J$.
Finally, we note that these results are in agreement with the analysis of the time-evolution of the electron density shown above.

We note here, that we confine our analysis about the interfacial transfer rate by choosing realistic experimental values for all parameters. 
However, introducing $m \to  \mu m$, $J \to \lambda J$ enables the scale transformation of the interfacial transfer rate $k_{Inj} \to k_{Inj}'$ into
\begin{equation} \label{scaling}
k_{Inj}'(\epsilon_{0}) = \mu^2 / \lambda \cdot k_{Inj}(\epsilon_{0}/\lambda)
\end{equation}
where $\mu$, $\lambda$ are the scaling constants of $m$, $J$ respectively. 
Through this scale transformation the results presented in Fig. \ref{Yield} can then be adapted to describe physically equivalent systems having a different scale $(k_{Inj}')$ and set of parameters ($\Omega'=\lambda\Omega,V'=\lambda V,g'=\lambda g,\sigma_{\epsilon}'=\lambda\sigma_{\epsilon}$). 

Finally, we remark the following: 
First, we have not found interfacial electron-transfer in the limit of extremely strong electron-phonon coupling ($\alpha^2 \gtrsim 8.0$).
This is readily explained since in this limit the coupling between the donor and the first site of the acceptor becomes much larger than the renormalized bandwidth of the polaronic sub-bands. 
Thus, one is, to a good extent, left with two eigenstates separated by a larger energy offset giving rise to charge localization at the interface. 

Second, we note that the results of our time-dependent study present some similarity with the Dirac-Frenkel time-dependent simulations in one-dimension \cite{bera2015impact} where their study has focused on the interplay between electron-vibration interaction and the interfacial Coulomb interaction between the hole and electron.
Yet, \cite{bera2015impact} has not taken spatial disorder into consideration and has focused on the quantum yield in absence of any electron-hole recombination. 
The quantum yield, in absence of recombination, then simply measures the charge injected at infinite distance from the interface, i.e. their quantum yield is simply $1 - P$ with $P$ being the weight of the wave-function localized near the interface. 
Instead, our study quantifies the charge injection-rate which is of more fundamental relevance since this quantity can be compared to the experimentally observed recombination rate.
The quantum yield, in the presence of recombination, can then be obtained by
\begin{equation}
Y \simeq \frac{ k_{Inj} }{ k_{Inj} + k_{Rec} } \, . 
\end{equation}
A high quantum yield is then given when $k_{Inj} >> k_{Rec}$ holds. 
Moreover, the applicability of the in Ref. \cite{bera2015impact} presented simulation is limited since their approximated treatment of the many-body nature of the polaronic state (momentum average approximation) is valid in the small bandwidth limit only. 
Contrary to Ref.\cite{bera2015impact}, the proposed I-DMFT approach is accurate over the entire parameter space \cite{richler2018inhomogeneous}.

\section{Conclusion} \label{Conclusion}
To summarize, we have applied the I-DMFT approximation to a generic one-dimensional model Hamiltonian, which parameters model the charge carrier dynamics in prototypical PCBM and \ce{C60} acceptor systems. 
Our results show that polaronic bands, when compared to spatial disorder, can provide here the main detrimental influence on the efficiency of charge transfer of electrons across organic interfaces. 
From this perspective, organic molecules with moderate reorganization energies should be used preferentially in next-generation materials since increasing the electron-phonon interaction hinders the range of suitable incoming electron energies due to the fragmentation of the local density of states into narrow polaronic sub-bands. 

Finally, we emphasize here that the easy numerical implementation of the I-DMFT approximation \cite{richler2018inhomogeneous} allows one to study a variety of recently proposed perhaps more realistic donor-acceptor model systems. 
In particular, I-DMFT enables to investigate the impact electric fields induced by energy level pinning \cite{liu2008control}, structural heterogeneity as a function of distance to the interface \cite{poelking2015design,mcmahon2011holes} and gradients in the energy landscape \cite{poelking2015impact,wilke2011electric}.
These problems were previously difficult to access but may assist the charge separation process drastically.
Open questions we reserve for future work. 

\section{Acknowledgments} 
The authors thank S. Fratini, S. Ciuchi and G. D'Avino for the stimulating discussions.
K.-D. Richler acknowledges the LANEF framework (ANR-10-LABX-51-01) for its support with mutualized infrastructure.

\appendix

\section{Krylov subspace method} \label{krylov_space}

We recall here a general definition of the in this work used Krylov subspace method. 
The Krylov subspace of dimension $N$ ($KS_N$) \cite{gutknecht2007brief} is the linear subspace spanned by:
\begin{equation}
KS_N (\ket{\phi_0}) = span \left \{ \ket{\phi_0}, H\ket{\phi_0}, ... , H^{N-1}\ket{\phi_0} \right \}.
\end{equation}
Here $\ket{\phi_0}$ represent a suitably chosen vector of the original Hilbert space and where we will denote in the following the orthonormalized basis vectors in $KS_N$ by
\begin{equation}
\ket{\mathcal{K}_{0}}, \ket{\mathcal{K}_{1}}, \ket{\mathcal{K}_{2}}, \dots, \ket{\mathcal{K}_{N-1}} \equiv KS_N \, 
\end{equation}
with $\ket{\mathcal{K}_{0}} = \ket{\phi_0}$ being defined as the initially chosen wave-function. 
An orthogonal basis of $KS_N$ can then be constructed with the method of Haydock \cite{Lanczos1950zz}, an iterative procedure that is capable of constructing the Krylov space via a three-term recurrence relation \cite{haydock1980solid}:
\begin{equation} \label{haydock recursion relations}
H \ket{\mathcal{K}_{n}} = a(n) \ket{\mathcal{K}_{n}} + b(n) \ket{\mathcal{K}_{n+1}} + b(n-1) \ket{\mathcal{K}_{n-1}}
\end{equation}
with the initial conditions $\ket{\mathcal{K}_{0}}=0$, $b(-1)=0$ and where $\ket{\mathcal{K}_{n}}$ obeys the orthogonality relation $\bra{\mathcal{K}_{n}}\ket{\mathcal{K}_{m}}=\delta_{n,m}$. 
The reduced Hamiltonian matrix $H_{ \mathcal{KS_N}}$ in $KS_N$ then reads
\begin{equation} \label{Haydock}
 H_{ \mathcal{KS_N}} = \left[ \begin{matrix}
a(0) & b(0) & 0 & 0  & \cdots & \\
b(0) & a(1) & b(1) & 0 & 0 & \cdots   \\
 0 &  b(1) & a(2)  & b(2) & 0  & \cdots \\
\vdots  & \ddots & \ddots & \ddots & \ddots & \ddots  \end{matrix}  \right] \, .
\end{equation} 
The time evolution of the wave function $\ket{\mathcal{K}_{0}}$ in the Krylov subspace $KS_N$ is then defined by 
\begin{eqnarray}
\ket{\psi(t)} &=& \text{e}^{-iH_{\mathcal{KS_N}}t} \ket{\mathcal{K}_{0}}  \\ \nonumber
&=& \sum^N_{l=0} \text{e}^{-iE_{\mathcal{KS_N}}(l)t} \bra{Z_{\mathcal{KS_N}}(l)}\ket{\mathcal{K}_{0}} \ket{Z_{\mathcal{KS_N}}(l)} \, . 
\end{eqnarray}
Here $E_{\mathcal{KS_N}}(l)$, $\ket{Z_{\mathcal{KS_N}}(l)}$ represent the l-th eigenvalue, respectively eigenvector of the Hamiltonian $H_{ \mathcal{KS_N}}$ which have been determined by exact diagonalization 
(due to greatly reduced size of the Hamiltonian matrix $H_{ \mathcal{KS_N}}$).
\begin{figure}[!b]
  \centering
 \includegraphics[scale=1.25]{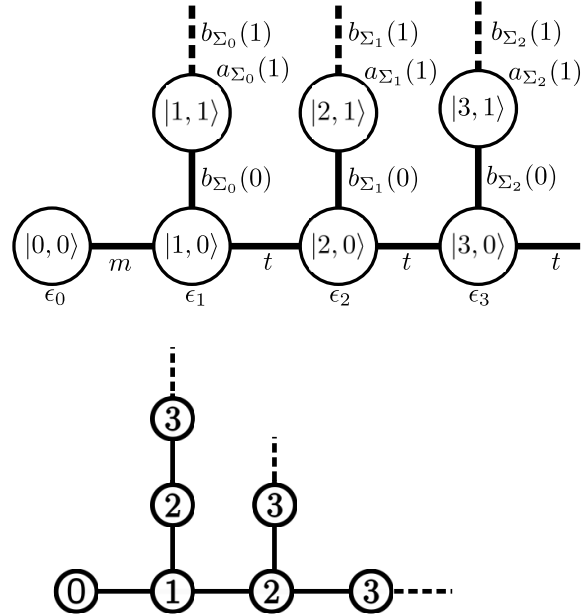}
    \caption{Top: tight-binding representation of the Hamiltonian $H_{IDMFT}$, where $H_{IDMFT}$ and, especially, the coefficients $a_{\Sigma_n}(n)$, $b_{\Sigma_n}(n)$ have been computed using the in Ref.\cite{richler2018inhomogeneous} presented I-DMFT formalism. Bottom: the most outward extension of the Krylov subspace vector $\ket{\mathcal{K}_{N}}$ for the first three steps of the recursion scheme.}
    \label{wave_function_krylov}
\end{figure} 

To further illustrate the extension of the Krylov subspace vector $\ket{\mathcal{K}_{N}}$, we have added in Fig.\ref{wave_function_krylov} the tight-binding representation of the Hamiltonian $H_{IDMFT}$, which has been computed with the in Ref.\cite{richler2018inhomogeneous} presented I-DMFT method, and which has been used to determine all  $\ket{\mathcal{K}_{N}}$. 
Starting Haydock's recursion from the initial state $\ket{\mathcal{K}_{0}}=\ket{\phi_0}=\ket{0,0}$ (electron at site $i=0$ having energy $\epsilon_0$ and no phonon modes excited), one then finds
\begin{eqnarray} \label{H_idmt_step_one}
H_{IDMFT} \ket{0,0} &=& m \ket{1,0} + \epsilon_0 \ket{0,0} \\ \nonumber
&=& a(0) \ket{\mathcal{K}_{0}} + b(0) \ket{\mathcal{K}_{1}}  \, . 
\end{eqnarray}
Here states in the energy independent tight-binding representation of $H_{IDMFT}$ are labeled $\ket{x,n}$ with $x$ being the lattice coordinate and $n$ being the phonon number (see Fig.\ref{wave_function_krylov}). 
Projecting on Eq.\ref{H_idmt_step_one} with $\ket{\mathcal{K}_{0}}$, respectively $\ket{\mathcal{K}_{1}}$ one then finds $a(0)=\epsilon_0$, $b(0)=m$, and $\ket{\mathcal{K}_{1}}=\ket{1,0}$.
In the second recursion step one then finds
\begin{eqnarray} 
H_{IDMFT} \ket{\mathcal{K}_{1}} &=& t \ket{2,0} + \epsilon_1 \ket{1,0} \\
&+& b_{\Sigma_0}(0)\ket{1,1} + m \ket{0,0}  \nonumber \\ 
&=& a(1) \ket{\mathcal{K}_{1}} + b(1) \ket{\mathcal{K}_{2}} + b(0) \ket{\mathcal{K}_{0}} \nonumber
\end{eqnarray}
The set of new recursion coefficients then are $a(1)=\epsilon_1$, $b(1)=\sqrt{t^2 + b^2_{\Sigma_0}(0)}$ and the new wave function reads
\begin{equation}
\ket{\mathcal{K}_{2}} = \frac{ t \ket{2,0} }{b(1)} + \frac{ b_{\Sigma_0}(0) \ket{1,1} }{b(1)} .
\end{equation}
This iterative three-term recurrence procedure is repeated until all recursion coefficients a(n), b(n) and Krylov subspace wave functions $\ket{\mathcal{K}_{N}}$ are determined. 

To conclude, apart from $\ket{\mathcal{K}_{0}}$, $\ket{\mathcal{K}_{1}}$, which correspond to a electron localized at site $i=0$, respectively $i=1$ with zero phonon modes excited, all other vectors $\ket{\mathcal{K}_{N}}$ represents excitations of the many-body system (electron and phonon modes) that progressively spread away from the interface into the acceptor (see Fig.\ref{wave_function_krylov}).

\nocite{}

\bibliographystyle{iopart-num}

\bibliography{kevin}

\end{document}